\documentclass[12pt]{article}
\usepackage{amsmath, amsthm, amssymb,amsfonts}

\usepackage{hyperref}
\usepackage[nosort]{cite}
\allowdisplaybreaks[1]

\begin{document}
\begin{titlepage}
\begin{center}
{\Large{
\centerline{\bf On the first law of entanglement for}
\centerline{\bf Quasi-Topological gravity}
}}
\vskip
0.5cm
{Shan-Quan Lan$^{a}$\footnote{shanquanlan@mail.bnu.edu.cn}, Gu-Qiang Li$^{a}$\footnote{zsgqli@hotmail.com}, Jie-Xiong Mo$^{a}$\footnote{mojiexiong@gmail.com}, Xiao-Bao Xu$^{a}$\footnote{xuxb@lingnan.edu.cn}} \vskip 0.05in
{\it ${}^a$ Institute of Theoretical Physics,\\ Lingnan Normal University, Zhanjiang, 524048 Guangdong, China }
\end{center}

\vskip 0.5in
\baselineskip 16pt
\abstract{The first law of entanglement has been used to obtain the linearized Einstein equations of the holographic dual spacetimes. In the present paper, the first law of entanglement in Quasi-Topological gravity is explicitly derived by using the Iyer-Wald formalism. In addition, we investigate the extended first law of entanglement for the special case in Quasi-Topological gravity.}
\end{titlepage}

\section{Introduction}
One of the most interesting questions in AdS/CFT~\cite{Maldacena:1997re,Gubser:1998bc,Witten:1998qj} is how the dynamical spacetime in AdS emerges from the CFT
physics. The past decade has seen considerable progress in addressing this problem~\cite{Ryu:2006bv,Hubeny:2007xt,Swingle:2009bg,VanRaamsdonk:2009ar,Czech:2012bh,
Lashkari:2013koa,Faulkner:2013ica,Almheiri:2014lwa,Pastawski:2015qua,Hayden:2016cfa,Dong:2016eik,Jafferis:2015del}, throughout which quantum entanglement plays an
essential role. It is well known since~\cite{Ryu:2006bv,Hubeny:2007xt} that entanglement entropy is holographic dual to the area of bulk extremal surfaces.
Motivated by this proposal, the authors of~\cite{Swingle:2009bg,VanRaamsdonk:2009ar} suggested that the structure of quantum entanglement in the QFT state
should play the crucial role in bulk reconstruction. Further evidences for the connection between geometry and entanglement, such as entanglement wedge
reconstruction, have appeared in \cite{Czech:2012bh,Almheiri:2014lwa,Pastawski:2015qua,Hayden:2016cfa,Dong:2016eik,Jafferis:2015del}.

At the same time, it was shown that the first law of entanglement for ball-shaped regions in the CFTs implies the Einstein equations linearized around the AdS
vacuum in the bulk~\cite{Lashkari:2013koa,Faulkner:2013ica}. The demonstration is based on the first law of entanglement, which exhibits a relation reminiscent
of the first law of thermodynamics~\cite{Blanco:2013joa}, i.e.
\begin{equation}
\delta S_A=\delta \langle H_A\rangle  \label{deltas}
\end{equation}
where $\delta S_A$ is the first variation of the entanglement entropy for a spatial region $A$, while $\delta \langle H_A\rangle$ is
the first-order variation in the expectation value of the modular Hamiltonian $H_A$. The last operator is defined as the logarithm of the unperturbed reduced
density matrix, i.e. $\rho_A\simeq e^{-H_A}$.

In general, there are few powerful tools to derive the linearized Einstein equations in the bulk from the first law of entanglement\footnote{Very recently,
with the replica trick, an approach to obtaining the integrated linearized Einstein equations around an arbitrary background from the first law of entanglement
is proposed in~\cite{Dong:2017xht}.}. However, in the case of a spherical entangling regions in the vacuum of a CFT the problem is much more
tractable~\cite{Casini:2011kv}. In particular, the modular Hamiltonian is given by a simple integral
\begin{equation}
H=2 \pi \int_{B(R,x_0)} d^{d-1}x\, \frac{R^2-|\vec{x}-\vec{x}_0|^2}{2 R}\ T_{00}(\vec{x}) \label{sphereH}
\end{equation}
where $T_{00}$ is the energy density of the CFT, and ${B(R,x_0)}$ denotes a ball of radius $R$ centered at $\vec{x}_0$ on a fixed time slice.
Thus, the first law eq.\eqref{deltas} for a ball-shaped region $B$ simplifies to
\begin{equation}
\delta S_B=\delta E_B  \label{deltas2}
\end{equation}
where $E_B$ is defined as
\begin{equation}
E_B \equiv 2 \pi \int_{B(R,x_0)} d^{d-1}x\, \frac{R^2-|\vec{x}-\vec{x}_0|^2}{2 R}\ \langle T_{00}(\vec{x})\rangle \label{eb}
\end{equation}
In order to analyze the linearized gravitational dynamics, we need to find a gravitational constraint in AdS which is holographic dual to the first law in the CFT, i.e.
\begin{equation}
\delta S_B^{grav}=\delta E_B^{grav}  \label{deltasg}
\end{equation}
The key ingredient in obtaining the gravitational constraint called the gravitational first law~\cite{Faulkner:2013ica} is a result that the vacuum density matrix
for $B$ maps to a thermal state of the CFT on hyperbolic space by a conformal transformation\footnote{We should mention that the fundamental role in deriving
the gravitational constraints is Ryu-Takayanagi prescription~\cite{Ryu:2006bv}, which is held up for the general entangling surfaces more than sphere.}, see more
details in~\cite{Casini:2011kv}. Moreover, it is found by~\cite{Faulkner:2013ica} that this gravitational first law  can be regarded as the AdS-Rindler analogue
of the Iyer-Wald first law for asymptotically flat black hole horizons. Then, by reversing the machinery used by Iyer and Wald \cite{Iyer:1994ys}to derive the first law from the
equations of motion, the linearized Einstein's equations denoted as $\delta \textbf{E}_g=0$ emerge from the gravitational first law.

In the last year, motivated by the extended black hole thermodynamics or black hole chemistry, the extended first law of entanglement in holography was
established~\cite{Caceres:2016xjz} including the variations of both the cosmological constant $\Lambda$ and the Newton's constant G. For Einstein gravity in
$AdS_3$, the extended first law of entanglement can be written as
\begin{equation}
\delta E=\delta S_{EE}-\frac{S_{EE}}{c}\delta c \label{efl}
\end{equation}
where $c$ is the central charge of two dimensional CFTs. When energy is fixed, from \eqref{efl} we see that the variation of entanglement entropy in CFT is
proportional to the change of central charge. Alternatively, it is easy to derive the relation
\begin{equation}
\delta S_{EE}=\frac{S_{EE}}{c}\delta c
\end{equation}
from the entanglement entropy of a single interval
\begin{equation}
S_{EE}=\frac{c}{3}\log\frac{l}{\epsilon}
\end{equation}
here $l$ is the length of the interval and $\epsilon$ is the UV regulator.
Hence the authors of \cite{Caceres:2016xjz} explicitly shows that entanglement entropy contains dynamic information about CFT in the gravity side of holography.
Similar results can be found in Lovelock gravity\cite{Kastor:2016bph}. Furthermore, the extended first law can give off-shell information about the bulk gravitational
theory \cite{Caceres:2016xjz}. Therefore it is interesting to investigate the extended first law of entanglement in other gravity theories, e.g. Quasi-Topological
gravity \cite{Myers:2010ru}, which includes cubic curvature interactions. Later, the studies of Quasi-Topological gravity with holography were exploited in
\cite{Myers:2010jv,Myers:2010xs,Hung:2014npa,Kuang:2010jc}.

In this note, we will restrict our attention to the gravitational first law of entanglement in Quasi-Topological gravity following \cite{Caceres:2016xjz}. For a spherical
entangling surface, the first law of entanglement in the CFT vacuum state was holographically derived within the dual Quasi-Topological gravity. It may be regarded
as a crucial ingredient in establishing the extended first law of entanglement in the CFT dual to Quasi-Topological gravity. We postpone a full analysis of the
extended first law of entanglement for Quasi-Topological gravity.

The paper is organized as follows. In section \ref{IW}, we briefly review the Iyer-Wald formalism which is essential to the gravitational first law
\eqref{deltasg}. In section \ref{QuasiTopo}, we introduce some facts about Quasi-Topological gravity, and then obtain the gravitational first law of entanglement along the
approach of \cite{Caceres:2016xjz}. Some results of the calculations are in the appendix \ref{detail}. Finally, in section 4 we briefly discuss possible generalizations.

\section{Overview of Iyer-Wald formalism}
\label{IW}
The Iyer-Wald formalism provides the most general derivation of the first law of black hole thermodynamics in terms of Noether charge \cite{Iyer:1994ys}.

Let $\mathbf{L}$ be a diffeomorphism invariant Lagrangian. It can written as a $(d+1)-$form:
\begin{equation}
\mathbf{L}=\cal L\,\varepsilon
\end{equation}
where $\varepsilon$ is the volume form
\begin{align}
\varepsilon&=\frac{1}{(d+1)!}\epsilon_{a_1\cdots a_{d+1}}dx^{a_1}\wedge\cdots\wedge dx^{a_{d+1}} \cr
&=\sqrt{-g}dt\wedge dx^1\cdots\wedge dx^d
\end{align}
For later convenience, we will also define:
\begin{align}
\varepsilon_a &=\frac{1}{d!}\epsilon_{a b_2\cdots b_{d+1}}dx^{b_2}\wedge\cdots\wedge dx^{b_{d+1}},\cr
\varepsilon_{ab} &=\frac{1}{(d-1)!}\epsilon_{a b c_3\cdots c_{d+1}}dx^{c_3}\wedge\cdots\wedge dx^{c_{d+1}}
\end{align}
Here $\epsilon$ is the Levi-Civita tensor, and our sign convention is $\epsilon_{t z x^1\cdots x^{d-1}}=+\sqrt{-g}$.

The dynamic fields in the gravitational theory are collectively denoted by $\phi=\left\{g_{\mu\nu},\ldots\right\}$. The variation of the Lagrangian has the following
form
\begin{equation}
\delta \mathbf{L}=\mathbf{E}^{\phi}\delta\phi+d\mathbf{\Theta}(\delta\phi)
\end{equation}
where $\mathbf{E}^\phi$ are the equations of motion for the theory, and $\mathbf{\Theta}$ is a boundary term called the symplectic potential current.

Let $\xi$ be an arbitrary vector field, the variation of the Lagrangian under a diffeomorphism generated by $\xi^\mu$ is $\delta_\xi \mathbf{L}=d(\xi\cdot\mathbf{L})$.
Then according to Noether's theorem, we can associate $\xi$ a current
\begin{equation}
\mathbf{J}[\xi]=\mathbf{\Theta}(\delta_\xi \phi)-\xi\cdot\mathbf{L}
\end{equation}
where the dot means the inner product of $\xi$ with the form $\mathbf{L}$. We can see that $\mathbf{J}$ is conserved on shell from
\begin{equation}
d\mathbf{J}[\xi]=-\mathbf{E}^\phi\delta_\xi\phi
\end{equation}
When the current $\mathbf{J}$ is conserved, we can have the Noether charge $\mathbf{Q}[\xi]$,
\begin{equation}
\mathbf{J}[\xi]=d\mathbf{Q}[\xi] \label{noethercharge}
\end{equation}
Next consider a variation $\delta \mathbf{J}$
\begin{align}
\delta \mathbf{J}[\xi]&=\delta \mathbf{\Theta}(\delta_\xi \phi)-\xi\cdot\delta \mathbf{L}\cr
&=\delta \mathbf{\Theta}(\delta_\xi \phi)-\xi\cdot d\mathbf{\Theta}(\delta \phi)\cr
&=\delta \mathbf{\Theta}(\delta_\xi \phi)-\delta_\xi \mathbf{\Theta}(\delta \phi)+d(\xi\cdot \mathbf{\Theta}(\delta \phi))
\end{align}
where we used the background equations of motion $E^\phi=0$ and the rules of the Lie derivative of a form,
\begin{equation}
\delta_\xi \mathbf{u}\equiv \mathcal L_\xi \mathbf{u}=\xi\cdot d\mathbf{u}+d(\xi\cdot\mathbf{u})
\end{equation}
If $\xi$ is a symmetry of all the fields, i.e. $\delta_\xi \phi=0$, then we have
\begin{equation}
\delta \mathbf{J}-d(\xi\cdot \mathbf{\Theta})=0
\end{equation}
When $\phi$ satisfies the equations of motion, using \eqref{noethercharge} the above equation can be written as
\begin{equation}
d(\delta \mathbf{Q}-\xi\cdot \mathbf{\Theta})=0 \label{differential first law}
\end{equation}
For a Cauchy surface with a boundary $\partial \Sigma$, integrating \eqref{differential first law} yields the following formula
\begin{equation}
\int_{\partial \Sigma}\chi=0\label{intfirstlaw}
\end{equation}
where $\chi$ is defined as
\begin{equation}
\chi=\delta \mathbf{Q}-\xi\cdot \mathbf{\Theta}\label{chi}
\end{equation}
To obtain the first law of black hole thermodynamics, we need choosing $\xi$ to be the timelike Killing vector which is null at the horizon and $\Sigma$ a time slice
of the black hole exterior. $\partial \Sigma$ have two parts, one at infinity and one at the horizon. The first law of black hole thermodynamics takes the form
\begin{equation}
\int_{\partial \Sigma_\infty}(\delta \mathbf{Q}-\xi\cdot\mathbf{\Theta})=\int_{\partial \Sigma_{horizon}}(\delta \mathbf{Q}-\xi\cdot\mathbf{\Theta}) \label{iyerwald}
\end{equation}
This equation is the same as the standard first law $\delta E=\frac{\kappa}{2\pi} \delta S$, because the integral at infinity is the variation in the canonical energy
$\delta E$, while the integral at the horizon is $\frac{\kappa}{2\pi} \delta S$. So we sketch the Iyer-Wald theorem.

To establish the gravitational first law of entanglement, we should resort the fact\cite{Casini:2011kv} that the entanglement entropy for a ball-shaped region $B$ in the CFT vacuum state is holographically
dual to the Wald entropy of the AdS-Rindler patch,
\begin{equation}
d s^2=-\frac{\rho^2-l^2}{R^2}d\tau^2+\frac{l^2 d\rho^2}{\rho^2-l^2}+\rho^2(du^2+\sinh^2 u\,d\Omega^2_{d-2}) \label{adsrindler}
\end{equation}
Therefore we should apply the Iyer-Wald theorem to the AdS-Rindler horizon.

In fact, according to \cite{Faulkner:2013ica}, for a ball-shaped region $B$ of radius $R$ centered at $x_0$ in the CFT,
the Ryu-Takayangi surface $\tilde{B}=\{t=t_0, (x^i-x^i_0)^2+z^2=R^2\}$ in Poincar\'e coordinates is the bifurcation surface of the Killing horizon for the Killing
vector $\xi_B$, which vanishes on $\tilde{B}$,
\begin{equation}
\xi_B=-\frac{2\pi}{R}(t-t_0)[z\partial_z+(x^i-x^i_0)\partial_i]+\frac{\pi}{R}[R^2-z^2-(t-t_0)^2-(\vec{x}-\vec{x_0})^2]\partial_t
\end{equation}
This vector is actually proportional to $\partial_\tau$ in the AdS-Rindler coordinates \eqref{adsrindler}.

Therefore the region $\Sigma$ enclosed by $\tilde{B}$ and $B$ is a spacelike slice which can be regarded as the hyperbolic 'black hole'(i.e. AdS-Rindler space) exterior.
Based on this result, we apply the Iyer-Wald theorem to the Cauchy surface $\Sigma$ and obtain the first law
\begin{equation}
\delta S^{Wald}_B=\delta E_B[\xi_B]
\end{equation}
where $\delta E_B[\xi_B]$ is the canonical energy appearing \eqref{iyerwald}. Moerover, we can show that $\delta E_B[\xi_B]$ is nothing but the "holographic"
energy dual to the modular energy \eqref{eb}, see \cite{Faulkner:2013ica}.

With these efforts, we derive out the gravitational first law of entanglement for an arbitrary higher-derivative gravity.

\section{The first law of entanglement in Quasi-\\
Topological gravity}
\label{QuasiTopo}
In this section, we derive the gravitational first law of entanglement for Quasi-Topological gravity.

Let us begin with the Lagrangian for $D=5$ Quasi-Topological gravity,
\begin{equation}
\mathcal L=\frac{R-2\Lambda}{16\pi G}+\alpha_2\mathcal{L}_{(2)}+\alpha_3\mathcal{L}_{(3)}
\end{equation}
where $\mathcal{L}_{(2)}$ is the four dimensional Euler density,
\begin{equation}
\mathcal L_{(2)}=R_{abcd}R^{abcd}-4R_{ab}R^{ab}+R^2
\end{equation}
and $\mathcal L_{(3)}$ is the curvature cubed interaction
\begin{align}
\mathcal L_{(3)}=&R_a{}^c{}_b{}^d R_c{}^e{}_d{}^f R_e{}^a{}_f{}^b+\frac{1}{56}(21R_{abcd}R^{abcd}R-72R_{abcd}R^{abc}{}_e R^{de}\cr
&+120R_{abcd}R^{ac}R^{bd}+144R_a{}^b R_b{}^c R_c{}^a-132R_a{}^b R_b{}^a R+15R^3) \label{cubicint}
\end{align}
Here, $\alpha_2=\frac{\lambda L^2}{32\pi G}$ is the Gauss-Bonnet coupling and $\alpha_3=\frac{7\mu L^4}{64\pi G}$ is the coupling of the curvature cubed interaction.

The action for Quasi-Topological gravity admits $AdS_5$ vacua as a solution\cite{Myers:2010ru}:
\begin{equation}
d s^2=\frac{\tilde{L}^2}{z^2}(dz^2-dt^2+d\vec x^2)
\end{equation}
where the radius of curvature of the $AdS_5$ spacetime is
\begin{equation}
\tilde{L}^2=\frac{L^2}{f_\infty} \label{scalecos}
\end{equation}
and the constant $f_\infty$ is one of the roots of
\begin{equation}
1-f_\infty+\lambda f_\infty^2+\mu f_\infty^3=0 \label{cubic}
\end{equation}
There are also the solutions describing planar AdS black holes
\begin{equation}
ds^2=\frac{r^2}{L^2}(-\frac{f(r)}{f_\infty}dt^2+dx_1^2+dx_2^2+dx_3^2)+\frac{L^2}{r^2f(r)}dr^2
\end{equation}
where $f(r)$ is given by roots of the following equation
\begin{equation}
1-f(r)+\lambda f(r)^2+\mu f(r)^3=\frac{r_0^4}{r^4}
\end{equation}
where $r_0$ is the radius of the horizon.

Besides these nice properties above, the most remarkable fact of Quasi-Topological gravity which will be used later is that the linearized equations of motion in
$AdS_5$ background precisely match the linearized equations in Einstein gravity. Combined with the known fact that the following perturbation
\begin{equation}
ds^2=\frac{\tilde{L}^2}{z^2}(dz^2+(-1+H z^4)dt^2+(1+\frac{H z^4}{3})(dx_1^2+dx_2^2+dx_3^2))\label{pertsolution}
\end{equation}
satisfies the linearized Einstein equations\cite{Caceres:2016xjz}, where $H$ is a small perturbation parameter, we will take the metric \eqref{pertsolution} as a on-shell
linear perturbation around $AdS_5$ vacuum in Quasi-Topological gravity.

Due to the higher derivative term, we can't use the Ryu-Takayanagi proposal \cite{Ryu:2006bv} to calculate the entanglement entropy of any region in the CFT. We should
adopt the entanglement entropy functional in higher derivative gravity theory proposed in \cite{Dong:2013qoa,Camps:2013zua}. For Quasi-Topological gravity, it has been
found \cite{Bhattacharyya:2014yga} that the surface in the bulk which minimizes the entanglement entropy functional for a ball-shaped region $B(R,0)$ in the CFT is
\begin{equation}
z=\sqrt{R^2-\vec{x}^2} \label{rt}
\end{equation}
where we choose $\vec{x}_0=0$ for simplicity. This result can also be inferred from the vanishing of the extrinsic curvature of the surface \eqref{rt}, which leads
the entanglement entropy functional reduce to the Wald functional of the horizon entropy \cite{Wald:1993nt}, then it is easy to see that the surface \eqref{rt} is a minimal surface
homologous to the boundary region $B(R,0)$ \cite{Faulkner:2013ica}.

So our set up is almost the same as the one in Gauss-Bonnet gravity \cite{Caceres:2016xjz}, of which the gravitational first law of entanglement for Gauss-Bonnet
gravity is obtained, see Eq. $(4.37)$ in \cite{Caceres:2016xjz}. To get the gravitational first law of entanglement for Quasi-Topological gravity, we only need to
focus on the cubic-curvature interaction $\mathcal L_3$, about which we will carry out some calculations.

Using the result presented in \cite{Bueno:2016ypa}, the symplectic potential current $\mathbf\Theta$ and the Noether charge $\mathbf Q$ were given by respectively:
\begin{align}
\mathbf \Theta&=\varepsilon_\mu(2P^{\mu\alpha\beta\nu}\nabla_\nu\delta g_{\alpha\beta}-2\nabla_\nu P^{\mu\alpha\beta\nu}\delta g_{\alpha\beta})\cr
\mathbf Q&=\varepsilon_{\mu\nu}(-P^{\mu\nu\rho\sigma}\nabla_\rho\xi_\sigma-2\xi_\rho\nabla_\sigma P^{\mu\nu\rho\sigma})         \label{currentcharge}
\end{align}
where $P^{\mu\nu\rho\sigma}$ is defined as
\begin{equation}
P^{\mu\nu\rho\sigma}=\frac{\partial \mathcal L}{\partial R_{\mu\nu\rho\sigma}}
\end{equation}
We now can use $\eqref{currentcharge}$ to check the equality $\eqref{iyerwald}$ for the cubic-curvature interaction $\eqref{cubicint}$. Firstly, we can show that
\begin{equation}
P^{\mu\nu\rho\sigma}=\frac{1}{16\pi G}\frac{9\mu}{2}(g^{\mu\nu}g^{\nu\sigma}-g^{\nu\rho}g^{\mu\sigma}) \label{P}
\end{equation}
for the cubic-curvature interaction $\eqref{cubicint}$ in pure $AdS_5$ background with the following identities
\begin{align}
&\frac{\partial R_{\mu\alpha\beta\nu}}{\partial R_{\sigma\rho\lambda\eta}}=\frac{1}{2}\big[\delta^{[\sigma}_\mu\delta^{\rho]}_\alpha\delta^{[\lambda}_\beta\delta^{\eta]}_\nu
+\delta^{[\lambda}_\mu\delta^{\eta]}_\alpha\delta^{[\sigma}_\beta\delta^{\rho]}_\nu\big],\quad
\frac{\partial R_{\rho\sigma}}{\partial R_{\mu\alpha\beta\nu}}=\delta^{[\alpha}_{(\rho}g^{\mu][\beta}\delta^{\nu]}_{\sigma)}\cr
&\frac{\partial R}{\partial R_{\mu\alpha\beta\nu}}=g^{\beta[\mu}g^{\alpha]\nu},\quad
\delta g^{\mu\nu}=-g^{\mu\alpha}g^{\nu\beta}\delta g_{\alpha\beta}
\end{align}
In fact, this calculation \eqref{P} supports the observation of \cite{Myers:2010tj}. Clearly, we have
\begin{equation}
\mathbf \Theta\propto\varepsilon_\mu(g^{\mu\beta}g^{\nu\alpha}-g^{\alpha\beta}g^{\mu\nu})\nabla_\nu\delta g_{\alpha\beta}
\end{equation}
From the result of the appendix A in \cite{Caceres:2016xjz}, we simply get
\begin{equation}
\mathbf \Theta=0
\end{equation}
for the cubic-curvature interaction $\eqref{cubicint}$.

After a straightforward calculation, we find the explicit formula of the Noether charge $\mathbf \Theta$ for the cubic-curvature interaction given in appendix A.
For the perturbative spacetime \eqref{pertsolution}, the variation of the Noether charge can be written as \cite{Caceres:2016xjz}
\begin{equation}
\delta \mathbf Q|_\Sigma=\sum_{i<j}\delta\mathbf Q_{zij}dz\wedge dx^i\wedge dx^j+\delta \mathbf Q_{123}dx^1\wedge dx^2\wedge dx^3
\end{equation}
We perform the calculation by using Mathematica and obtain
\begin{align}
\delta\mathbf Q_{zij}&=\alpha_3 H\frac{176\pi z}{7\tilde{L} R}\varepsilon_{ijk}x^k\cr
\delta\mathbf Q_{123}&=\alpha_3 H\frac{24(2\pi r^2+11\pi z^2-2\pi\vec{x}^2)}{7\tilde{L} R}
\end{align}
So the form of $\chi$ \eqref{chi} on the bifurcation surface \eqref{rt} is given by
\begin{equation}
\chi|_{\partial\Sigma_h}=\alpha_3\frac{8\pi(39R^2-61\vec{x}^2)}{7\tilde{L} R}H dx^1\wedge dx^2\wedge dx^3
\end{equation}
Then we have
\begin{equation}
\int_{\partial\Sigma_h}\chi=\alpha_3\frac{128\pi^2 R^4}{35\tilde{L}}H\label{convhorizon}
\end{equation}
Meanwhile, as we have mentioned, the entanglement entropy of the ball-shaped region $B(R,0)$ in the CFT can be computed by the Wald functional
\cite{Wald:1993nt}
\begin{equation}
S_{Wald}=-2\pi\int_{\mathcal H} d^n x\sqrt{h}\frac{\partial \mathcal L}{\partial R_{abcd}}\hat{\varepsilon}_{ab}\hat{\varepsilon}_{cd}\label{waldfunctinal}
\end{equation}
where $\hat{\varepsilon}_{ab}$ is the binormal to the horizon.

Using the formula (6.4) of \cite{Faulkner:2013ica}, the binormal $\hat{\varepsilon}_{ab}$ in the perturbative spacetime \eqref{pertsolution} take the form
\begin{align}
\hat{\varepsilon}_{tz}&=-\frac{\tilde{L}^2}{Rz}(1-\frac{H z^4}{2})(1+\frac{H z^4\vec{x}^2}{6R^2})\cr
\hat{\varepsilon}_{ti}&=-\frac{\tilde{L}^2 x^i}{Rz^2}(1-\frac{H z^4}{2})(1+\frac{H z^4\vec{x}^2}{6R^2})
\end{align}
Now the Wald functional \eqref{waldfunctinal} can be written as
\begin{equation}
S_{Wald}=-2\pi\int_{\mathcal H} d^n x\sqrt{h}\big(4\frac{\partial \mathcal L}{\partial R_{tztz}}\hat{\varepsilon}_{tz}\hat{\varepsilon}_{tz}
+8\frac{\partial \mathcal L}{\partial R_{tzti}}\hat{\varepsilon}_{tz}\hat{\varepsilon}_{ti}
+4\frac{\partial \mathcal L}{\partial R_{titi}}\hat{\varepsilon}_{ti}\hat{\varepsilon}_{ti}\big)
\end{equation}
Then we can show that
\begin{align}
&\frac{\partial \mathcal L}{\partial R_{tzti}}|_{pert}=0\cr
&\frac{\partial \mathcal L}{\partial R_{tztz}}|_{pert}=\frac{18z^4(-3+Hz^4)^4}{7\tilde{L}^8(-1+Hz^4)(3+Hz^4)^4}\cr
&\frac{\partial \mathcal L}{\partial R_{titi}}|_{pert}=\frac{54z^4(-3+Hz^4)^2(9+Hz^4(26+Hz^4))}{7\tilde{L}^8(-1+Hz^4)(3+Hz^4)^5}
\end{align}
Thus we find that the variation of the entanglement entropy for the spherical entangling surface is
\begin{equation}
\delta S_{EE}=-\alpha_3\frac{128 \pi^2R^4}{35\tilde{L}}H\label{vee}
\end{equation}
Combing \eqref{convhorizon} and \eqref{vee}, we have
\begin{equation}
\int_{\partial \Sigma_h}\chi=-\delta S_{EE}
\end{equation}
This is one of the expected formulas. For the integral of $\chi$ on the boundary $B(R,0)$, we have
\begin{equation}
\int_{\partial \Sigma_\infty}\chi=\alpha_3\frac{128 \pi^2R^4}{35\tilde{L}}H
\end{equation}
Therefore we established the first law for Quasi-Topological gravity theory by using the Iyer-Wald formalism,
\begin{equation}
\delta S_{EE}=\delta E
\end{equation}

\section{The extended first law for Quasi-Topological gravity}
In this section we find that there could be an extended first law for Quasi-Topological gravity which is analogous to holographic entanglement chemistry in Gauss-Bonnet gravity. In particular, using the extension of the Iyer-Wald formalism \cite{Caceres:2016xjz} we get an extended first law for Quasi-Topological gravity with varying the coupling of curvature interaction at fixed the cosmological constant $\Lambda$ and Newton's constant $G$. We will focus on five-dimensional Quasi-Topological gravity, because it is straightforward to generalize the result obtained here to Quasi-Topological gravity in $d+1$ dimensions.

To get the correct extended first law with varying the coupling of curvature interaction, we need an analytic expression of the cosmological constant $\Lambda$ in terms of $\tilde{L}$, $G$, $\alpha_2$ and $\alpha_3$. Specifically, the cosmological constant $\Lambda$ is a solution of the equation \eqref{scalecos}
\begin{equation}
\tilde{L}^2=L^2/f_\infty
\end{equation}
where $L^2=-\frac{d(d-1)}{2\Lambda}$, we will take $d=4$.

However, we find that there is no such analytic function for the cosmological constant $\Lambda$ when the values of $\alpha_2$ and $\alpha_3$ are general, due to $f_\infty$ is a complicated function of $\Lambda$ as a solution of the cubic equation \eqref{cubic}, i.e. $1-f_\infty+\lambda f_\infty^2+\mu f_\infty^3=0$.
We also note that $\alpha_2=\frac{\lambda L^2}{32\pi G}$ and $\alpha_3=\frac{7\mu L^4}{64\pi G}$ for five-dimensional Quasi-Topological gravity. From the analysis of \cite{Myers:2010ru}, the cubic equation \eqref{cubic} is equivalent to the following equation
\begin{equation}
x^3-3px-2q=0      \label{cubic1}
\end{equation}
by the relation $f_\infty=x-\frac{\lambda}{3\mu}$, and we defined that
\begin{equation}
p=\frac{3\mu+\lambda^2}{9\mu^2} \quad\quad\quad\quad q=-\frac{2\lambda^3+9\mu\lambda+27\mu^2}{54\mu^3}
\end{equation}
The cubic equation \eqref{cubic1} will reduce to a perfect cubic equation when $p=0$, i.e. $\mu=-\frac{\lambda^2}{3}$. We will consider this special case and assume $\lambda<0$, then
\begin{equation}
f_\infty=(2q)^\frac{1}{3}+\frac{1}{\lambda}=(\frac{3}{\lambda^2}-\frac{1}{\lambda^3})^\frac{1}{3}+\frac{1}{\lambda}>0
\end{equation}
So the cosmological constant $\Lambda$ is
\begin{equation}
\Lambda=-\frac{2(3\tilde{L}^4-96G\tilde{L}^2\pi\alpha+1024G^2\pi^2\alpha^2)}{\tilde{L}^6}
\end{equation}
where $\alpha=\alpha_2=\frac{\lambda L^2}{32\pi G}$. With $f_\infty>0$, we have $h'(f_\infty)=-1+2\lambda f_\infty+3\mu f_\infty^2<0$. So this $AdS$ vacua is ghost-free. From the equation
\begin{equation}
x^3-2q_r=0
\end{equation}
where $q_r=-\frac{2\lambda^3+9\mu\lambda+27\mu^2(1-\omega^4/r^4)}{54\mu^3}$, this $AdS$ vacua can support can a black hole with a horizon at $r=\omega$.\\
In the following, the Lagrangian is conveniently written as
\begin{equation}
\mathcal L=\frac{R-2\Lambda}{16\pi G}+\alpha \mathcal{L}_{(2)}-\frac{112}{3}G\pi\alpha^2 \mathcal{L}_{(3)}
\end{equation}
where $\alpha\equiv\alpha_2$.\\
Then we will find the extended first law of entanglement with varying $\alpha$ and $G$ and $L$ fixed. Because the metric doesn't have the explicitly dependence
on $\alpha$, we have:
\begin{equation}
\delta g_{\mu\nu}=0 \label{persol}
\end{equation}
From the equation of motion for Quasi-Topological gravity, it is easy to see that \eqref{persol} is a solution to the extended linearized equation of motion. The extended first law with $\delta \alpha$ takes the form:
\begin{equation}
\delta \alpha\int_\Sigma\frac{\partial \mathcal{L}}{\partial \alpha}\xi\cdot\varepsilon-\int_{\partial \Sigma_\infty}\chi+\int_{\partial \Sigma_h}\chi=0
\label{gfl}
\end{equation}
with
\begin{align}
\frac{\partial \mathcal{L}}{\partial \alpha}&=-\frac{1}{8\pi G}\frac{\partial \Lambda}{\partial \alpha}+\mathcal L_{(2)}-\frac{224}{3}\pi G\alpha \mathcal L_{(3)}\cr
&=-\frac{1}{8\pi G}(\frac{192G\pi}{\tilde{L}^4}-\frac{4096(G\pi)^2}{\tilde{L}^6}\alpha)+\mathcal L_{(2)}-\frac{224}{3}\pi G\alpha \mathcal L_{(3)}
\end{align}
In \cite{Caceres:2016xjz}, the authors have derived the extended first law of entanglement with varying $\alpha$ in Gauss-Bonnet theory. So in this paper,
we will only consider the modification to the extended first law of entanglement by the term $\mathcal L_{(3)}$ in the Largaragian of Quasi-Topological gravity. \\
The variation of the Neother charge $\delta Q$ can be found from \eqref{charge} in the appendix:
\begin{equation}
\delta Q=-\frac{224}{3}\pi G\alpha\delta \alpha\varepsilon_{\mu\nu}\{(\cdots)\nabla_\rho\xi_\sigma-2\xi_\rho(\cdots)\}
\end{equation}
For a pure AdS, it is trivially that $\xi_\rho(\cdots)=0$, and $\delta Q$ is reduced to
\begin{equation}
\delta Q=-\frac{224}{3}\pi G\alpha\delta \alpha\varepsilon_{\mu\nu}\nabla_\rho\xi(\cdots)
\end{equation}
Next, according to the fact that $\delta g_{\mu\nu}=0$ means the symplectic potential current vanishes, we have
\begin{equation}
\Theta=0
\end{equation}
Therefore the Iyer-Wald form coincides with $\delta Q$. Specifically, the Iyer-Wald form is as follows
\begin{align}
\chi&=\delta Q=\frac{224}{3}G\pi\alpha\delta \alpha\varepsilon_{\mu\nu}\nabla_\rho\xi_\sigma(g^{\mu\rho}g^{\nu\sigma}-g^{\mu\sigma}g^{\nu\rho})\frac{18}{7}\frac{1}{\tilde{L}^4}\cr
&=768G\pi\alpha\delta \alpha\frac{1}{\tilde{L}^4}[\varepsilon_{ti}\frac{2\pi z^2x^i}{R\tilde{L}^2}+\varepsilon_{tz}\frac{z^2}{\tilde{L}^2}(\frac{2\pi z}{R}+\frac{\xi^t(t=0)}{z})]
\end{align}
The restriction of $\chi$ to the horizon is
\begin{equation}
\chi|_{\partial \Sigma_h}=\frac{1536\,G\pi^2 R\alpha\,\delta\alpha}{\tilde{L}(R^2-\vec{x}^2)^2}dx^1\wedge dx^2\wedge dx^3
\end{equation}
By calculating the integral of $\chi$ over the horizon, we get
\begin{equation}
\int_{\partial \Sigma_h}\chi=\frac{1536\,G\pi^2 R\alpha\,\delta\alpha}{\tilde{L}}{\rm Vol}(S^2) \int_0^{\sqrt{R^2-\epsilon^2}}\frac{r^2}{(R^2-r^2)^2}dr\label{chihorizon}
\end{equation}
Furthermore, we can compute $\delta S_{EE}$ for the cubic term $\mathcal L_{(3)}$ directly by differentiating the unperturbed entanglement entropy
\eqref{waldfunctinal} with respect to $\alpha$. The result agrees with the above one \eqref{chihorizon}
\begin{equation}
\int_{\partial \Sigma_h}\chi=-\delta^{(\alpha)} S_{EE}\label{gflh}
\end{equation}
On the other hand, the restriction of $\chi$ to the boundary is
\begin{equation}
\chi|_{\partial \Sigma_\infty}=768G\pi^2\alpha\delta\alpha\frac{1}{\tilde{L}R}(\frac{1}{\epsilon^2}+\frac{R^2-\vec{x}^2}{\epsilon^4})dx^1\wedge dx^2\wedge dx^3
\end{equation}
Integrating over the boundary, we have
\begin{equation}
\int_{\partial \Sigma_\infty}\chi=768G\pi^2\alpha\delta\alpha\frac{1}{\tilde{L}R}{\rm Vol}(S^2)\int_0^{\sqrt{R^2-\epsilon^2}}(\frac{1}{\epsilon^2}+\frac{R^2-r^2}{\epsilon^4})r^2dr\label{gflb}
\end{equation}
At last, we calculate the first integral in \eqref{gfl}
\begin{equation}
\delta\alpha\int_\Sigma\frac{\partial \mathcal{L}}{\partial \alpha}\xi\cdot\varepsilon=\frac{768G\pi^2\alpha\delta \alpha}{\tilde{L}R}{\rm Vol}(S^2)\int_0^{\sqrt{R^2-\epsilon^2}}(\frac{1}{R^2-r^2}+\frac{R^2-r^2}{\epsilon^4}-\frac{2}{\epsilon^2})r^2dr\label{dalpha}
\end{equation}
Then we can use \eqref{gflh}, \eqref{gflb}, \eqref{dalpha} and the identity\cite{Caceres:2016xjz}
\begin{equation}
\int_0^{\sqrt{R^2-\epsilon^2}}(\frac{3}{\epsilon^2}-\frac{1}{R^2-r^2})r^2dr=2R^2\int_0^{\sqrt{R^2-\epsilon^2}}\frac{r^2}{(R^2-r^2)^2}dr
\end{equation}
to recast the extended first law of entanglement as
\begin{equation}
-\frac{1536G\pi^2R\alpha\delta\alpha}{\tilde{L}}{\rm Vol}(S^2)\int_0^{\sqrt{R^2-\epsilon^2}}\frac{r^2}{(R^2-r^2)^2}dr-\delta S_{EE}=0
\end{equation}
We note that this is just a part of the extended first law of entanglement for Quasi-Topological gravity. In order to get the completed one, we need also taking account of the extended first law of entanglement for Gauss-Bonnet gravity\cite{Caceres:2016xjz}. Combining these result, we have
\begin{equation}
(\frac{24\pi\tilde{L}}{R}+\frac{768G\pi^2\alpha}{\tilde{L}R})\delta \alpha(-2R^2){\rm Vol}(S^2)\int_0^{\sqrt{R^2-\epsilon^2}}\frac{r^2}{(R^2-r^2)^2}dr-\delta S_{EE}=0
\end{equation}
With the expression of entanglement entropy for Quasi-Topological gravity, we find
\begin{equation}
(\frac{24\pi\tilde{L}}{R}+\frac{768G\pi^2\alpha}{\tilde{L}R})\delta \alpha(-2R^2)\frac{S_{EE}}{\frac{R\tilde{L}^3}{4G}(1-6\times\frac{32\pi G\alpha}{\tilde{L}^2}-48\times\frac{64G^2\pi^2\alpha^2}{\tilde{L}^4})}-\delta S_{EE}=0
\end{equation}
Hence, the extended first law is
\begin{equation}
\delta^{(\alpha)}S_{EE}=\Psi_{\alpha}\delta \alpha
\end{equation}
where
\begin{equation}
\Psi_{\alpha}=-\frac{192G\pi(\tilde{L}^2+32G\pi\alpha)}{(\tilde{L}^4-192G\pi\alpha\tilde{L}^2-3072G^2\pi^2\alpha^2)}S_{EE}
\end{equation}

\section{Conclusion and outlook}
In this paper, we derived the first law of entanglement for a ball-shaped region in the CFT whose holographic dual is Quasi-Topological gravity theory.
We found this relation mainly using the Iyer-Wald formalism. This result suggests that Quasi-Topological gravity has well-defined properties for the linearized gravitational perturbation. It will be interesting to use other properties of entanglement entropy to find more constraints on Quasi-Topological gravity theory, for example the positivity of relative entropy \cite{Banerjee:2014oaa}. We also investigate the extended first law of entanglement in Quasi-Topological gravity. Because there is no an analytic function of the cosmological constant $\Lambda$ for the general $\alpha_2$ and   $\alpha_3$, we only discussed the case $\mu=-\frac{\lambda^2}{3}$ in which the cosmological constant $\Lambda$ has an exact form. Although we just consider the $AdS_5$ case, it is straightforward to obtain the similar result in $AdS_{d+1}$.

Another interesting problem is developing the first law of entanglement and its gravity dual for d=4 $\rm U(1)$ gauge theory, in which Maxwell theory is conformal. To get the entanglement entropy of Maxwell theory, we need to include the entanglement entropy of edge modes, which are the boundary degrees of freedom localized on the entangling surface generated by the gauge redundancy \cite{Donnelly:2014fua}. This result is equivalent to the statement that the contact term in geometric entropy is the entanglement entropy of the edge modes. Moreover, for d=4 Maxwell theory, Donnelly and Wall \cite{Donnelly:2014fua} find that the inclusion of the edge modes resolves the discrepancy between conformal anomaly and entanglement entropy in the universal logarithmic term noticed by \cite{Dowker:2010bu}. One difficulty in formulating the first law of entanglement for d=4 $\rm U(1)$ gauge theory is what the boundary terms in the modular Hamiltonian corresponding to the edge modes are. On the other hand, the authors of \cite{Donnelly:2015hxa} argue that the Ryu-Takayanagi entanglement entropy already includes an edge mode contribution, see also \cite{Harlow:2016vwg,Lin:2017uzr}, so it is desirable to find the deep connection between the entanglement entropy of edge modes and Iyer-Wald formalism, we refer to \cite{Donnelly:2016auv,Speranza:2017gxd} for the related discussions.

\section*{Acknowledgements}
We would like to thank Phuc H. Nguyen for helpful correspondence. The research of X. B. Xu is supported by the Lingnan Normal University Project ZL1501.
The work of G. Q. Li is supported by Natural Science Foundation of Guangdong Province, China, under Grant Nos. 2016A030307051 and 2015A030313789.
The work of J. X. Mo is supported by NSFC grants (No.11605082) and Natural Science Foundation of Guangdong Province, China, under Grant No. 2016A030310363.

\appendix
\section{the Noether charge for the cubic-curvature interaction}
\label{detail}
The $P^{\mu\nu\rho\sigma}$ for the cubic-curvature interaction $\mathcal L_{(3)}$ is
\begin{align}
P^{\mu\nu\rho\sigma}_{(3)}&=\alpha_3 \biggl\{\frac{3}{4}(R^{\nu e\sigma f}R^\mu{}_e{}^\rho{}_f-R^{\nu e\rho f}R^\mu{}_e{}^\sigma{}_f-\mu\leftrightarrow\nu)\cr
&+\frac{1}{56}\bigg(21[2R^{\mu\nu\rho\sigma} R+\frac{1}{2}R_{abcd}R^{abcd}(g^{\mu\rho}g^{\nu\sigma}-g^{\mu\sigma}g^{\nu\rho})]\cr
&-18[2(R^{\mu\nu\rho}{}_e R^{\sigma e}-R^{\mu\nu\sigma}{}_e R^{\rho e}+R^{\rho\sigma\mu}{}_e R^{\nu e}-R^{\rho\sigma\nu}{}_e R^{\mu e})\cr
&+(R_{abc}{}^\nu R^{abc\sigma}g^{\mu\rho}-R_{abc}{}^\nu R^{abc\rho}g^{\mu\sigma}-\mu\leftrightarrow\nu)]\cr
&+60[(R^{\mu\rho}R^{\nu\sigma}-R^{\mu\sigma}R^{\nu\rho})+(R^\nu{}_b{}^\sigma{}_d R^{bd}g^{\mu\rho}-R^\nu{}_b{}^\rho{}_d R^{bd}g^{\mu\sigma}-
\mu\leftrightarrow\nu)]\cr
&+108[g^{\mu\rho}R^{\sigma c}R_c{}^\nu-g^{\mu\sigma}R^{\rho c}R_c{}^\nu-\mu\leftrightarrow\nu]\cr
&-66[(g^{\mu\rho}R^{\nu\sigma}R-g^{\mu\sigma}R^{\nu\rho}R-\mu\leftrightarrow\nu)+R_{ab}R^{ab}(g^{\mu\rho}g^{\nu\sigma}-g^{\mu\sigma}g^{\nu\rho})]\cr
&+\frac{45}{2}R^2(g^{\mu\rho}g^{\nu\sigma}-g^{\mu\sigma}g^{\nu\rho})\bigg)\biggr\}\label{generalP}
\end{align}

Evaluating the Noether charge $Q_\xi$ associated with the Killing vector $\xi$ by \eqref{generalP}£¬ we find
\begin{align}
\mathbf Q&=\alpha_3\varepsilon_{\mu\nu}\biggl\{-\biggl\{\frac{3}{4}(R^{\nu e\sigma f}R^\mu{}_e{}^\rho{}_f-R^{\nu e\rho f}R^\mu{}_e{}^\sigma{}_f
-\mu\leftrightarrow\nu)\cr
&+\frac{1}{56}\bigg(21[2R^{\mu\nu\rho\sigma} R+\frac{1}{2}R_{abcd}R^{abcd}(g^{\mu\rho}g^{\nu\sigma}-g^{\mu\sigma}g^{\nu\rho})]\cr
&-18[2(R^{\mu\nu\rho}{}_e R^{\sigma e}-R^{\mu\nu\sigma}{}_e R^{\rho e}+R^{\rho\sigma\mu}{}_e R^{\nu e}-R^{\rho\sigma\nu}{}_e R^{\mu e})\cr
&+(R_{abc}{}^\nu R^{abc\sigma}g^{\mu\rho}-R_{abc}{}^\nu R^{abc\rho}g^{\mu\sigma}-\mu\leftrightarrow\nu)]\cr
&+60[(R^{\mu\rho}R^{\nu\sigma}-R^{\mu\sigma}R^{\nu\rho})+(R^\nu{}_b{}^\sigma{}_d R^{bd}g^{\mu\rho}-R^\nu{}_b{}^\rho{}_d R^{bd}g^{\mu\sigma}-
\mu\leftrightarrow\nu)]\cr
&+108[g^{\mu\rho}R^{\sigma c}R_c{}^\nu-g^{\mu\sigma}R^{\rho c}R_c{}^\nu-\mu\leftrightarrow\nu]\cr
&-66[(g^{\mu\rho}R^{\nu\sigma}R-g^{\mu\sigma}R^{\nu\rho}R-\mu\leftrightarrow\nu)+R_{ab}R^{ab}(g^{\mu\rho}g^{\nu\sigma}-g^{\mu\sigma}g^{\nu\rho})]\cr
&+\frac{45}{2}R^2(g^{\mu\rho}g^{\nu\sigma}-g^{\mu\sigma}g^{\nu\rho})\bigg)\biggr\}\nabla_\rho\xi_\sigma\cr
&-2\xi_\rho\biggl\{\frac{3}{2}(2\nabla^{[\nu}R^{e]f}R^\mu{}_e{}^\rho{}_f+R^{\nu e\sigma f}\nabla_\sigma R^\mu{}_e{}^\rho{}_f-\mu\leftrightarrow\nu)\cr
&+\frac{1}{56}\bigg(21[-4\nabla^{[\mu}R^{\nu]\rho}R+2R^{\mu\nu\rho\sigma}\nabla_\sigma R+(\nabla_\sigma R_{abcd})R^{abcd}(g^{\mu\rho}g^{\nu\sigma}-
g^{\mu\sigma}g^{\nu\rho})]\cr
&-18[2(\nabla_\sigma R^{\mu\nu\rho}{}_e R^{\sigma e}+\frac{1}{2}R^{\mu\nu\rho}{}_e\nabla^e R-2\nabla^{[\mu}R^{\nu]e}R_e{}^\rho-
R^{\mu\nu\sigma}{}_e\nabla_\sigma R^{\rho e}\cr
&-2\nabla^{[\mu}R^{e]\rho}R^\nu{}_e+R^{\rho\sigma\mu}{}_e\nabla_\sigma R^{\nu e}+2\nabla^{[\nu}R^{e]\rho}R^\mu{}_e-
R^{\rho\sigma\nu}{}_e\nabla_\sigma R^{\mu e})\cr
&+(\nabla_\sigma R_{abc}{}^\nu R^{abc\sigma}g^{\mu\rho}-2R_{abc}{}^\nu\nabla^{[a}R^{b]c}g^{\mu\rho}-\nabla_\sigma R_{abc}{}^\nu R^{abc\rho}g^{\mu\sigma}-
R_{abc}{}^\nu \nabla_\sigma R^{abc\rho}g^{\mu\sigma}-\mu\leftrightarrow\nu)]\cr
&+60[(\nabla_\sigma R^{\mu\rho}R^{\nu\sigma}+\frac{1}{2}R^{\mu\rho}\nabla^\nu R-\mu\leftrightarrow\nu)\cr
&+(2\nabla^{[\nu}R^{b]d}R_{bd}g^{\mu\rho}+R^{\nu b\sigma d}\nabla_\sigma R_{bd}g^{\mu\rho}-\nabla^\mu R^{\nu b\rho d}R_{bd}-R^{\nu b\rho d}\nabla^\mu R_{bd}
-\mu\leftrightarrow\nu)]\cr
&+108[\frac{1}{2}g^{\mu\rho}\nabla^c R R_c{}^\nu+g^{\mu\rho}R^{\sigma c}\nabla_\sigma R_c{}^\nu-\nabla^\mu R^{\rho c}R_c{}^\nu-R^{\rho c}\nabla^\mu R_c{}^\nu
-\mu\leftrightarrow\nu]\cr
&-66[(\frac{1}{2}g^{\mu\rho}\nabla^\nu R R+g^{\mu\rho}R^{\nu\sigma}\nabla_\sigma R-\nabla^\mu R^{\nu\rho}R-R^{\nu\rho}\nabla^\mu R-\mu\leftrightarrow\nu)\cr
&+2\nabla_\sigma R_{ab}R^{ab}(g^{\mu\rho}g^{\nu\sigma}-g^{\mu\sigma}g^{\nu\rho})]\cr
&+45\nabla_\sigma R R(g^{\mu\rho}g^{\nu\sigma}-g^{\mu\sigma}g^{\nu\rho})\bigg)
\cr
&\biggr\}\biggr\}\label{charge}
\end{align}

\bibliographystyle{JHEP}
\bibliography{i}
\end{document}